\documentclass[prd, aps, superscriptaddress, preprintnumbers, twocolumn, floatfix, nofootinbib]{revtex4}

\usepackage{amsfonts}
\usepackage{amsmath}
\usepackage{amssymb}
\usepackage{bm}
\usepackage{dcolumn}
\usepackage{graphicx}   
\usepackage[latin1]{inputenc}
\usepackage{latexsym}
\usepackage{rotating}
\usepackage{hyperref}
\usepackage{graphicx}
\usepackage{color}

\usepackage{amsfonts}
\usepackage{amsmath}
\usepackage{amssymb}
\usepackage{bm}
\usepackage{dcolumn}
\usepackage{graphicx}
\usepackage[latin1]{inputenc}
\usepackage{latexsym}
\usepackage{rotating}
\usepackage{hyperref}

\newcommand\be{\begin{equation}}
\newcommand\ba{\begin{eqnarray}}
\newcommand\ee{\end{equation}}
\newcommand\ea{\end{eqnarray}}

\begin{document}

\title {On the Possible Enhancement of the Global $21$-cm Signal at Reionization from the Decay of Cosmic String Cusps}

\author{Robert Brandenberger}
\email{rhb@physics.mcgill.ca}
\affiliation{Physics Department, McGill University, Montreal, QC, H3A 2T8, Canada}

\author{Bryce Cyr}
\email{bryce.cyr@mail.mcgill.ca}
\affiliation{Physics Department, McGill University, Montreal, QC, H3A 2T8, Canada}

\author{Timoth\'ee Schaeffer}
\email{timothee.schaeffer@polytechnique.edu}
\affiliation{Ecole Polytechnique, Saclay, 91128 Palaiseau, France and \\
Physics Department, McGill University, Montreal, QC, H3A 2T8, Canada}

\date{\today}

\begin{abstract}

We consider cosmic string cusp annihilations as a possible source of enhancement to the global background radiation temperature in $21$-cm photons at reionization. A soft photon spectrum is induced via the Bremsstrahlung and Synchrotron emission of electrons borne out of QCD jets formed off the cusp. The maximal energy density background comes from synchrotron induced photons with a string tension of $G\mu \sim 10^{-18}$. In this instance, the radiation background at reionization is heated up by $5\cdot 10^{-3} \,\, K$. We find that the depth of the absorption trough ($\delta T_b$) in $21$-cm at reionization is altered by one part in $10^4$ from the strings, requiring high precision measurements to be detectable. This mechanism cannot explain the $\delta T_b$ observed by the EDGES experiment.

\end{abstract}

\pacs{98.80.Cq}
\maketitle

\section{Introduction} 

Cosmic strings are one dimensional topological defects that can form in the early universe. They form if the universe undergoes a phase transition in which the true vacuum state is degenerate, and not simply connected. If they form, the string region has a small, but macroscopic, width, and is made up of scalar and gauge particles related to the previously unbroken symmetry group (see e.g. \cite{CSrevs} for some reviews). Detection of these strings would provide invaluable insights into the mathematical structure of matter in the very early universe, and could be an alternative to expensive particle physics experiments when studying physics at these very high energies (see e.g. \cite{RHBrev}). The theory of cosmic strings is an elegant one to work with, particularly because they possess only one free parameter, the string tension $\mu$. This parameter is often expressed in a dimensionless way as $G\mu$, where $G$ is Newton's gravitational constant.

The strings themselves have been proposed as candidates for many unsolved mysteries in astrophysics and cosmology. Historically they were first thought of as a competition with inflation for providing the seeds of the large scale structure, though this required a string tension of $G\mu \sim 10^{-6}$ \cite{stringsLSS}. In contrast to inflation, a cosmological model in which the fluctuations are due to strings exclusively does not give rise to acoustic oscillations in the angular power spectrum of the cosmic microwave background (CMB) \cite{noacoustic}. Detailed studies of the acoustic peak structure of the angular power spectrum of the CMB now limits the string tension to values $G\mu < 10^{-7}$ \cite{CMBconstraints}. Since strings predict \cite{KS} lines in the sky across which the temperature of the CMB jumps, searches for these non-Gaussian features of string maps have the potential of reducing this upper bound by a couple of orders of magnitude \cite{Danos}. Even if strings are only a subdominant mechanism for structure formation, they leave behind distinctive non-Gaussian localized signals in the distribution of structure in the universe. Long string segments produce wakes \cite{wakes}, planar overdensities, while string loops seed compact clumps which correspond to nonlinear density fluctuations at early times. These nonlinear seed fluctuations could yield an explanation for the origin of globular clusters \cite{GC}, fast radio bursts \cite{bryceFRB, superconductingFRB}, and they could seed supermassive black holes \cite{SMBH}. Provided that the distribution of strings includes a scaling distribution of string loops (which is indicated by certain numerical studies \cite{CSsimuls}), pulsar timing arrays could provide a stronger constraint, $G\mu < 10^{-11}$ coming from  the observational upper bound on the amplitude of the stochastic gravitational wave background \cite{PTAconstraints}. 

The energy scale $\eta$ of particle physics models which yields cosmic strings could, however, be many orders of magnitude lower than that which yields a value of $G\mu$ corresponding to the upper bounds mentioned above. The non-observation of new physics beyond the particle physics Standard Model at the Large Hadron Collider sets a lower bound on $G \mu$ of the order of $10^{-30}$ (making use of the relation $\mu \simeq \eta^2$). As we will review below, the number of cosmic string loops at any given time increases as $G\mu$ decreases. Hence, it is possible that searching for effects of cosmic strings in the sky might lead to a lower bound on the cosmic string tension. In this work we will consider possible constraints on the cosmic string tension arising from an observational upper bound on the excess photon radiation at redshifted $21{\rm{cm}}$ wavelenghts. We find that the strongest signal arises from very low string tensions of $G\mu \sim 10^{-18}$,
but that the predicted signal lies below the current observational bounds.

Recently there has been some excitement regarding an anomalous radio background detected by the ARCADE-2 experiment \cite{ARCADE2}. This experiment detected a new radiation field that dominates over the CMB at low energies ($\nu < 1\,\,GHz$). The effective frequency dependent temperature of the background radiation was fitted to a power law \cite{ARCADE2Gil}
\begin{align}
T(\nu) \, = \, T_{CMB} + \xi T_R \left(\frac{\nu}{\nu_0} \right)^{\beta} \, ,
\end{align}
with $T_R = 1.19 \,\,K$, $\nu_0=1\,\, GHz$, and a spectral index of $\beta = -2.62$ \footnote{$T_{CMB}$ is the temperature of the CMB.}. Since the detected signal hasn't been fully disentangled from the background, the parameter $\xi$ sets the fraction of the excess radiation that was present at early times. This signal is not currently explained by radio point sources, galactic emission, or CMB spectral distortions.

The inclusion of a new radiation field present at early times can have a substantial effect on the $21$-cm signal predicted at reionization. When neutral hydrogen undergoes the ``forbidden'' hyperfine transition between its triplet and singlet state, it emits a $21$-cm photon. Conversely, CMB photons can be absorbed by cold clouds of neutral hydrogen by exciting hydrogen molecules to their excited state (see e.g. \cite{Furl} for a review). Thus, detection of these photons provides an accurate way to probe the distribution of neutral hydrogen in the universe. Recently, the first detection of a global (i.e. averaged over the sky) $21$-cm signal from the epoch of reionization was reported by the EDGES experiment \cite{EDGES}. The global signal detected was larger than what was expected according to the standard cosmological paradigm. While met with some scepticism about its validity \cite{EDGEScritic}, we are entering a time when this signal will likely be probed by other collaborations. What is certain is that the global $21{\rm{cm}}$ signal is not larger than the reported value from the EDGES collaboration. 

In this work we will compute the contribution of cosmic string loops to the global $21{\rm{cm}}$ signal \footnote{See \cite{Oscar} for recent work computing the contribution of cosmic string wakes to this signal.}, and we investigate whether, using the value reported by the EDGES collaboration as an upper bound for a possible signal, constraints on the cosmic string tension can be derived.

The observable measured for such a signal is the so-called differential brightness temperature, 
\begin{align}
\delta T_b(\nu) \, = \, T_b(\nu) -T_{CMB} \, ,
\end{align}
where the brightness temperature at the frequency $\nu$ is 
\begin{align}
T_b(\nu) \, \approx \, I_{\nu} c^2/2k_B\nu^2 \, ,
\end{align}
where  $I_{\nu}$ is the radiation intensity at frequency $\nu$, $c$ is the speed of light and $k_B$ is Boltzmann's constant \footnote{In the following we shall use natural units in which $c$, $k_B$ and Planck's constant are set to one.}. The differential brightness temperature is
\begin{align}
\delta T_b \, \propto \, 1-\frac{T_{\gamma}}{T_{spin}} \, ,
\end{align}
where $T_{\gamma}$ is the temperature of the photon radiation, and $T_{spin}$ is the spin temperature of the hydrogen, a weighted average of the CMB temperature, kinetic temperature from collisions with other particles, and scattering by Lyman $\alpha$ photons. Interestingly, the EDGES collaboration detected a value of $\delta T_b$ in absorption that was roughly twice what was predicted by the standard cosmological paradigm.

To explain this result, a flurry of papers were written to either lower the spin temperature (for example, by coupling to milli-charged dark matter as a way to sap energy from the hydrogen gas \cite{milliDark}), or increase the background radiation temperature. Feng and Holder showed that if $\mathcal{O}(1 \%)$ of the radio background detected by ARCADE-2 was present at reionization, the detected brightness temperature could be achieved \cite{ARCADE2Gil}.

If a distribution of cosmic string loops form, they will naturally persist throughout most of our cosmic history. These loops possess mechanisms that allow them to source electromagnetic radiation. We compute the induced background of $21$-cm photons present at reionization by this mechanism, in an attempt to source any part of this anomalous radio background.

In the following section, we review the {\it cusp annihilation} mechanism by which non-superconducting strings yield bursts of particles including photons. In Section III we compute the flux of radio frequency photons produced by the distribution of cosmic string loops present between the time $t_{rec}$ of recombination and the time of reionization (the time relevant for the global $21{\rm{cm}}$ signal). In Section IV we then infer the contribution which cosmic string loops make to the global $21{\rm{cm}}$ signal. 

\section{Cusp Emission Mechanism}

If a phase transition in the very early universe permits the production of cosmic strings, a network of strings will inevitably form \cite{Kibble}. The network of strings consists of both long strings (strings with curvature radius which is comparable or larger than the Hubble radius $t$) and loops. Due to their relativistic motion, the long string network coarsens with time by splitting off string loops. By causality \cite{Kibble}, the string network has to  persist until the current time. As can be shown using analytical arguments \cite{CSrevs}, the long string network approaches a {\it scaling solution} for which the curvature radius of the strings tracks the Hubble radius $t$ at all times. There are good reasons to believe that the distribution of string loops will also approach a scaling solution \cite{scalingSolution} according to which the statistical properties of the loop distribution are independent of time is all lengths are scaled to $t$. Numerical simulations \cite{CSsimuls} of the evolution of cosmic string networks using the Nambu-Goto action (i.e. neglecting the finite width of the strings) confirm that such a scaling solution is achieved, although field theory simulations \cite{Hindmarsh} still yield a different result (in these simulations it is the classical scalar and gauge field equations which are evolved).  In this work, we will assume a scaling distribution of string loops.

By considering the Nambu-Goto action for a string, one can derive its equations of motion. From these equations of motion, it follows \cite{KibbleTurok} that for each string loop with radius $R$, there will be at least one cusp per loop oscillation time $R^{-1}$. A cusp is a point on the string where the string velocity reaches the speed of light, and the string doubles back on itself. Since the string has a finite width, at each cusp point there will be an interval of length $l_c$ on the string where the two string segments (the parts of the string on either side of the cusp point) overlap. This region is  called the cusp. Whereas a long straight string is protected against decaying by topology, the cusp region looks like a string and an anti-string (strings with opposite winding numbers). It is not protected against decay,  and so this overlapping region self-annihilates (in analogy to particle - antiparticle annihilation), producing a jet of gauge and scalar particles that make up the string \cite{cuspPartProduction}. Previous work has been done on high energy gamma ray and neutrino signals from these annihilations \cite{robertGamma} \cite{robertNeutrino}, but these used an incorrect parametrization of the cusp length.

By considering special relativistic effects, one can show that the proper parametrization for the cusp yields the following form \cite{olumCusp} for the cusp length
\begin{align}
l_c \, \sim \, R^{1/2} w^{1/2}
\end{align}
where $w \sim \mu^{-1/2}$ is the microscopic width of the string \cite{widthRef}, and $\mu$ is the mass per unit length of the string. A consequence of this parametrization discussed in \cite{olumCusp} is that we get $\mathcal{O}(1)$ particles produced per cusp annihilation. Particles produced at the cusp are thought to decay to superheavy fermions with mass $Q_f$, before transferring their energy into a jet of standard model particles. 
The jets thus formed are similar to those produced at the LHC, containing pions, photons, etc. Previous works have considered this idea \cite{robertGamma}, and by using the Pion multiplicity function, it is possible to show that the low energy ($E_{\pi} << Q_f$) spectrum of neutral and charged pions emanating from a cusp is
\begin{align} \label{dist}
\frac{dN}{dE} \, = \, \frac{15}{24}\frac{\mu l_c}{Q_f^2}\left(\frac{Q_f}{E}\right)^{3/2}
\end{align}

The dominant pion decay channels are
\begin{align*}
\pi^0 &\rightarrow 2\gamma\\
\pi^+ &\rightarrow \mu^+ + \nu_{\mu}\\
\pi^- &\rightarrow \mu^- + \bar{\nu}_{\mu}
\end{align*}
And so photons with energies $E_{\gamma} > m_{\pi}$ will be sourced by neutral pion decay and follow the above spectrum, scaling with $E^{-3/2}$. However, we are interested in radio photons which are {\it soft} in comparison to the pion mass. For these photons we cannot apply the above scaling of the photon flux as a function of energy.  Whereas there will be primordial soft photons from jets, their spectrum is not known (at least to us). Hence, we will consider more robust production mechanisms for these soft photons.

We consider the decays of the charged pions into muons (and quickly thereafter, electrons). Assuming no further interactions of the electrons after production, they traverse the universe from the time of their production until today. As they travel, they interact with electric and magnetic fields, which source the production of photons through Bremsstrahlung and Synchrotron emission. The pion-produced electrons will once again follow the same spectrum above for $E_{e} > m_{\pi}$. From this distribution, we closely follow \cite{blumenthal} for computing the Bremsstrahlung and Synchrotron photon spectra, which can safely be extrapolated down to $21$-cm energies.

\section{Global $21$-cm Signal at Reionization from Cusp Annihilations}

\subsection{Photons Produced Via Bremsstrahlung}

First, let us compute the soft photon spectrum at reionization produced by electrons via the Bremsstrahlung process. The distribution of electrons produced by cusp annihilations at any one point in time is given by integrating (\ref{dist}) over all of the loops present. Loops form at a time $t$ with a characteristic radius $R=\alpha t$, and an analysis of the decay of these loops by gravitational radiation \cite{CSgrav} gives a cutoff radius $R_c = \gamma G\mu t$, below which loops live less than one Hubble expansion time and for which the energy density is negligible. Here, $\alpha \sim 1$, $\gamma \sim 10$ are dimensionless numerical constants fit by simulation \cite{CSsimuls}. For electrons emitted off of cusps at a time $t''$, their energy distribution is given by
\begin{align} \label{electrons}
\frac{dn_e (t'')}{dE(t'') dt''} \, = \,  \int_{\gamma G\mu t''}^{\alpha t''} n_{R, loops}(t'') \frac{dN}{dE} \frac{1}{R} dR \, ,
\end{align}
where the factor of $1/R$ comes from the fact that cusp annihilations occur on average once every oscillation time, $1/R$, and $n_{R, loops}$ is the number density of loops. 

The loop distribution follows a scaling solution which for $R > \gamma G\mu t$ is given by \cite{CSrevs}
\begin{equation}
  n_{R,loops} (t) =\begin{cases}
    \nu R^{-5/2} t^{-3/2} \,\,\,\,\,\,\,\,\,\,\,\,\,\,\,\,\, t< t_{eq}\\
    \nu R^{-5/2} t_{eq}^{1/2} t^{-2} \,\,\,\,\,\,\,\,\,\,\,\, t> t_{eq} \,\,\,\,\,\, t_f < t_{eq}\\
    \nu R^{-2} t^{-2} \,\,\,\,\,\,\,\,\,\,\,\,\,\,\,\,\,\,\,\,\,\,\,\,\,\,\, t> t_{eq} \,\,\,\,\,\, t_f > t_{eq}
  \end{cases}
\end{equation}
where $\nu$ is a coefficient of the order one \footnote{This constant has nothing to do with the frequency $\nu$ used earlier in the text.} which is determined from numerical simulations \cite{CSsimuls}, and $t_{eq}$ is the time of equal matter and radiation. We take $Q_f \sim \eta$, the symmetry breaking scale at which the strings formed (note also that $\eta \simeq \mu^{1/2}$). The integral over $R$ is dominated by the lower cutoff radius $\gamma G\mu t$. For values of $G\mu$ smaller than the current upper bound, loops of such radii are produced in the radiation phase. On the other hand, we only consider the contribution of electrons after the time of recombination to the global $21{\rm{cm}}$ signal. Hence, when  evaluating the integral (\ref{electrons}), we must use the expression for the loop distribution given in the second line of the above equation. Hence, the dominant piece of this integral yields
\begin{align}
\frac{dn_e (t'')}{dE(t'')dt''} \, \sim \, \nu \gamma^{-2} (G\mu)^{-3/2} t_{eq}^{1/2} t''^{-4}\cdot E(t'')^{-3/2} G^{-1/2} \, .
\end{align}
This expression gives us the energy spectrum of electrons emitted from cusps at time $t''$.

The next step is to determine the spectrum of electrons present at time $t'$ coming from electrons emitted from cusps at earlier times $t''$. To obtain this, we integrate over time from recombination to some $t' > t_{rec}$ to determine the electron distribution from all cusp annihilations over that timeframe. We have to take into account both the redshifting of the electron number density and the Jacobian of the energy ratio. By noting that the Jacobian transformation between energies at different times is
\begin{align}
E^{3/2}(t'')\frac{dn_e(t'')}{dE(t'')} \, = \,  E^{3/2}(t') \frac{dn_e(t')}{dE(t')} \left(\frac{t'}{t''}\right)^{1/3} \left(\frac{t'}{t''}\right)^2
\end{align}
and including the redshifting of the number density of the electrons, we find
\begin{eqnarray}
\frac{dn_e(t')}{dE(t')} &=& \nu \gamma^{-2} (G\mu)^{-3/2} t_{eq}^{1/2} E(t')^{-3/2} G^{-1/2} \nonumber \\
& & \, \times \int_{t_{rec}}^{t'} dt'' \left(\frac{t''}{t'}\right)^{1/3} \left(\frac{t''}{t'}\right)^2 t''^{-4}\\
& \sim & \, \nu \gamma^{-2} (G\mu)^{-3/2}  G^{-1/2} E(t')^{-3/2}t_{eq}^{1/2} t'^{-7/3}t_{rec}^{-2/3} \nonumber
\end{eqnarray}

Now that we have a time dependent expression for the electron distribution, we can compute the induced soft photon spectrum. Following \cite{blumenthal}, in the weak shielding limit the induced photon distribution per unit time is
\begin{align} \label{result2}
\frac{dn_{\gamma}(t')}{dE(t')dt'} \, \approx \, \left(\frac{8}{3}m_{\pi}^{-1/2}\right)\alpha_{em}r_0^2 K_{e}(t') E(t')^{-1} \sum_s n_s(t') \tilde{\phi}_w \, ,
\end{align}
where $\alpha_{em}$ is the EM fine structure constant, $r_0$ is the reduced Compton wavelength of the electrons, $r_0 = m_e^{-1}$, $n_s$ is the number density of a species of charged particles that our induced electrons interact with (we include a summation over different types of scatterers), and $\tilde{\phi}_w \sim 300$ is a dimensionless weak shielding coefficient (its value is taken from that of neutral hydrogen). This expression is for a power law of electrons going as $E^{-3/2}$, and $K_e$ is set by the expression $dn_e(t')/dE(t') = K_e(t') E(t')^{-3/2}$ \footnote{To obtain this expression, we consider the spectrum of photons produced by a single high energy electron, and then integrate over the distribution of electrons computed above.}.

To obtain the photon spectrum at the time $t$ when the global $21{\rm{cm}}$ signal is computed, we must integrate the above result (\ref{result2}) in time over all times when photons can travel without scattering, i.e. from $t_{rec}$ to time $t$. In computing this integral, we must take into account the Jacobian arising from converting $E(t')$ to $E(t)$ which is
\begin{align}
E(t')\frac{dn_{\gamma}(t')}{dE(t')} \, = \, E(t) \frac{dn_{\gamma}(t)}{dE(t)} \left(\frac{t}{t'}\right)^{2} \, 
\end{align}
as well as the redshifting of the number density of photons. Integrating over time and including the abovementioned effects, the spectrum of soft photons by Bremsstrahlung emission is
\begin{eqnarray}
\frac{dn_{\gamma}(t)}{dE(t)} &\approx&  \, \nu \gamma^{-2}(G\mu)^{-3/2} G^{-1/2}(\Lambda \tilde{\phi}_w)E(t)^{-1}t_{eq}^{1/2} t_{rec}^{-2/3} \nonumber \\ 
& & \times\int_{t_{rec}}^t \left( n_s(t)\left( \frac{t}{t'}\right)^2\right) \left(\frac{t'}{t}\right)^2 t'^{-7/3} dt'\\
&\sim & \, \nu \gamma^{-2}(G\mu)^{-3/2}G^{-1/2}(\Lambda \tilde{\phi}_w n_s(t))E(t)^{-1} t_{eq}^{1/2} t_{rec}^{-2}
\, , \nonumber
\end{eqnarray}
where
\begin{equation}
\Lambda \, = \, \frac{8}{3}\alpha_{em}r_0^2m_{\pi}^{-1/2} \, .
\end{equation}
Note that the integral is dominated by the earliest times, i.e. $t_{rec}$. Note that the amplitude of the spectrum increases as $G\mu$ decreases, the reason being that the increase in the number of loops as $G\mu$ decreases is a more important factor than the decrease in the cusp energy of an individual loop \footnote{This phenomenon was already seen in the computations of the high energy gamma and neutrino fluxes from string loops.}.

 At first glance it appears that as the value of the string tension decreases one would get a large and observable signal. However, as was shown in \cite{robertSpectral}, there is a crossover value of $G\mu$ below which gravitational radiation is not the dominant decay mechanism, and cusp annihilation becomes the more efficient method to lose energy. Assuming that cusp radiation dominates over gravitational radiation, one obtains an effective cutoff radius in the loop distribution which at time $t$ is 
\begin{equation}
 R_c^{cusp} \, = \, \left( \frac{3}{2} \mu^{-1/4} \beta^{-1} t\right)^{2/3} \, ,
\end{equation} 
with $\beta \sim 10$ being a parameter that measures how circular the loops are on average (perfectly circular loops have $\beta = 2\pi$).  If $R_c^{cusp}$ is larger than the gravitational radiation cutoff radius $\gamma G\mu t$ the cusp annihilation is the dominant decay process.

For values of $G\mu$ below the value where the gravitational cutoff radius equals $R_c^{cusp}$ at the time of recombination (the time which dominates in the integrals we perform), the computation of the soft photon distribution has to be redone, replacing the gravitational cutoff radius by $R_c^{cusp}$.This yields a soft photon spectrum of
\begin{eqnarray}
\frac{dn_{\gamma}^{cusp}(t)}{dE} &\approx&  \nu \beta^{4/3} (G\mu)^{5/6} G^{-5/6} E(t)^{-1} (\Lambda \tilde{\phi}_w n_s(t)) \nonumber \\
& & \, \times t_{eq}^{1/2} \int_{t_{rec}}^t dt' t'^{-7/3} \ln \left(\frac{t'}{t_{rec}}\right) \, .
\end{eqnarray}
In this region of values of $G\mu$, the signal decreases as $\mu$ decreases. Hence, we see that the largest soft photon signal is produced by values of $G\mu$ where the gravitational radiation cutoff equals $R_c^{cusp}$ at $t_{rec}$.

By equating the cutoff radii in the two regimes, the crossover value of the string tension can be derived. This value depends on time $t_c$ and is
\begin{align}
G\mu \, = \, \left(\left(\frac{2}{3}\right)^{2/3} \gamma \beta^{2/3} G^{-1/6} t_c^{1/3} \right)^{-6/7} \, .
\end{align}
Equivalently, we can fix $G\mu$, and in this case for times before $t_c$ cusp annihilation is the dominant decay mechanism, and for times after, gravitational radiation takes over. Taking our crossover time to be recombination, we find that $\frac{dn_{\gamma}(t)}{dE}$ is the proper expression for $G\mu > 10^{-18}$, and for smaller values of the tension we must use $\frac{dn_{\gamma}^{cusp}(t)}{dE}$.

\subsection{Photons Produced Via Synchrotron}

Next, we consider the soft photon produced by Synchrotron emission over the same timeframe. The computation of $dn_e(t')/dE(t')$ proceeds in the same way as for the Bremsstrahlung computation. From \cite{blumenthal}, the induced Synchrotron spectrum from a power law distribution of electrons, from recombination to reionization, is
\begin{eqnarray}
\frac{dn_{\gamma}(t')}{dE(t')dt'} \, &=& \,  \frac{2D_e(t') e^3}{m_e}B(t')^{(p+1)/2} \left(\frac{3e}{4\pi m_e}\right)^{(p-1)/2} \nonumber \\
& & \, \times a(p) E(t')^{-(p+1)/2} (2\pi)^{(p-1)/2} \, ,
\end{eqnarray}
where $B(t')$ is the magnetic field at time $t'$, $a(p)$ is a dimensionless constant of order $10^{-1}$, and $D_e(t')$ is defined by the equation
\begin{align}
\frac{dn_e(t')}{d\Gamma'} \, = \,  \frac{D_e(t')}{4\pi} \Gamma'^{-p} \, ,
\end{align}
where $\Gamma' = E(t')/m_e$.  For our case, $D_e(t')$ is
\begin{align}
D_e(t') \, = \,  4\pi\gamma^{-2}(G\mu)^{-3/2}G^{-1/2}m_e^{-1/2}t_{eq}^{1/2}t_{rec}^{-2/3}t'^{-7/3} \, .
\end{align}
As well, the power law is unchanged for the electrons so $p=3/2$. In deriving this expression, it was assumed that the cusp was an isotropic emitter of electrons. If a more detailed analysis is required, one must take into account the fact that the QCD jets coming off the cusp are beamed into a solid angle related to the strong coupling constant \cite{robertGamma}.

To obtain the photon spectrum at some late time $t$, we must integrate over the rate of production at earlier times $t'$, taking into accouint the redshifting of the number density, and of the Jacobian transformation of the energies, which for this power law integral is
\begin{align}
E^{5/4}(t')  \frac{dn_{\gamma}(t')}{dE'} \, = \,  E^{5/4}(t) \frac{dn_{\gamma}(t)}{dE(t)} \left(\frac{t}{t'}\right)^{1/6} \left(\frac{t}{t'}\right)^2 \, .
\end{align}
The computation yields a photon spectrum of
\begin{eqnarray}
\frac{dn_{\gamma}(t)}{dE(t)} \, &=& \, \Omega  E^{-5/4}(t)(G\mu)^{-3/2}t_{eq}^{1/2} t_{rec}^{-2/3} \nonumber \\
& & \, \times \int_{t_{rec}}^{t} dt' \left(B(t_{rec}) \left(\frac{t_{rec}}{t'}\right)^{4/3}\right)^{5/4} \nonumber \\
& & \, \times \left(\frac{t'}{t}\right)^{1/6} \left(\frac{t'}{t}\right)^2 t'^{-7/3} \\
&=& \,  \Omega (G\mu)^{-3/2} E^{-5/4}(t) B^{5/4}(t_{rec}) t_{eq}^{1/2} t_{rec}^{1/6} t^{-13/6} \, , \nonumber
\end{eqnarray}
with
\begin{equation}
\Omega \, = \, 4\pi\frac{2e^3}{m_e^{3/2}} \left(\frac{3e}{2 m_e}\right)^{1/4} a\left(\frac{3}{2}\right)  \gamma^{-2} G^{-1/2} \, .
\end{equation}
Since we intend to use the upper bound on the magnitude of the $B$ field from CMB observations, we have replaced $B(t')$ by $B(t_{rec})$, taking into account the appropriate redshift factor. Similarly to the Bremsstrahlung analysis, this spectrum is only valid for $10^{-18} < G\mu$ when gravitational radiation is the dominant decay mechanism. For smaller values of $G\mu$, we again use the cutoff radius for cusp decay, and find
\begin{eqnarray}
\frac{dn_{\gamma}^{cusp}(t)}{dE} \, &=& \,  \Omega'  E(t)^{-5/4} B^{5/4}(t_{rec}) (G\mu)^{5/6} t_{eq}^{1/2} \nonumber \\
& & \, \times \int_{t_{rec}}^{t} dt'  \left(\frac{t_{rec}}{t'}\right)^{5/3} \left(\frac{t'}{t}\right)^{13/6} t'^{-7/3} \ln \left(\frac{t'}{t_{rec}}\right) \nonumber \\
&=& \, \Omega'(G\mu)^{5/6} E^{-5/4}(t) B^{5/4}(t_{rec}) t_{eq}^{1/2} t_{rec}^{5/3} t^{-13/6} \nonumber \\
& & \, \times \int_{t_{rec}}^t dt' t'^{-11/6} \ln \left(\frac{t'}{t_{rec}}\right) \, ,
\end{eqnarray}
with
\begin{equation}
\Omega' \, = \, 4\pi \frac{2e^3}{m_e^{3/2}}\left(\frac{3e}{2m_e}\right)^{1/4} a\left(\frac{3}{2}\right) \beta^{4/3}  G^{-5/6}
\end{equation}
The above expressions must be evaluated and compared to the energy density in the CMB at $21$-cm wavelengths to infer any possible effects.

\section{Comparison with the CMB}

The first measurements of the global $21$-cm signal at reionization have started to come back \cite{EDGES}, and they seem to be in tension with current theory computations. In addition, the ARCADE-2 experiment has detected an anomalous radio background that may have some cosmological origin \cite{ARCADE2}. The global $21$-cm signal is characterized by an absorption depth ($\delta T$), and a shape. The signal reported by the EDGES experiment indicates that this depth may be twice as deep as was expected. The depth goes as $\delta T \propto T_r/T_{spin}$, where $T_r$ is the background radiation temperature at that time. Hence, a larger depth of the signal could be explained either by decreasing the spin temperature or by increasing the photon temperature. Usually, the radiation temperature $T_r$ is taken to be $T_r = T_{CMB}$. However, if there is extra production of soft photons then $T_r$ can be boosted. As we have seen, cosmic string cusp annihilation leads to a flux of soft photons and hence increases the effective radiation temperature at $21{\rm{cm}}$ frequencies. In this section we will compute the magnitude of the resulting absorption feature.

The energy density at $21$-cm in the CMB at reionization is given in natural units by
\begin{eqnarray}
\frac{d\rho_{CMB}}{dE} \, &=& \, \frac{8\pi h}{c^3}\frac{E^3}{e^{hE / kT}-1} \nonumber \\
&\approx& \, E^2 T \nonumber \\
&\approx& \,  10^{-40} \,\, GeV^3
\end{eqnarray}
when computed at a temperature of $T_{reion}=50\,\,K$. We wish to compute the ratio of the energy density from our string induced soft photon spectra, to that in the CMB. We define this ratio as
\begin{align}
\mathcal{F} \, = \, \frac{d\rho_{21-cm}/dE}{d\rho_{CMB}/dE} 
\end{align}
where the energy density in $21$-cm at reionization can be evaluated by computing $E_{21}\frac{dn_{\gamma}(t_{reion})}{dE}$ with $E_{21} \sim 10^{-15} \,\, GeV$. 

In the Bremsstrahlung analysis, the scatterers which contribute are the ionized particles which are present after recombination. The ionization fraction $f$ drops from $f = 1$ close to recombination to $f \sim 10^{-4}$ at later times (but before reionization) \cite{Holder}. We hence use
\begin{equation}
n_s(t) \, = \, f \rho_B(t) m_B^{-1} \, ,
\end{equation}
where $\rho_B$ is the baryon density and $m_B$ is the baryon (hydrogen) mass. The ratio $\mathcal{F}$ then becomes
\begin{eqnarray}
\mathcal{F} \, &\sim& \, \nu \gamma^{-2} (G\mu)^{-3/2} f \frac{m_{pl}}{m_B} \bigl( \frac{m_{pl}}{m_{\pi}} \bigr)^{1/2} \bigl( \frac{T_{rec}}{E} \bigr)^2 \bigl( \frac{r_0}{t_0} \bigr)^2 \nonumber \\
& & \, \times \frac{t^{2/3} t_{eq}^{1/3}}{t_{rec}} \bigl( \frac{t_0}{t} \bigr)^2 \, ,
\end{eqnarray}
where $t_0$ is the present time.

With this, the ratio in energy densities in $21$-cm at reionization (for interesting values of $G\mu$) is
\begin{eqnarray}
\mathcal{F}_{Brem} \, &\sim& \, 10^{-26} \,\,\,\,\,\,\,\,\,\,\,\,\,\,\,\,\, \textrm{For} \,\, G\mu = 10^{-7}\\
\mathcal{F}_{Brem} \, &\sim& \, 10^{-10} \,\,\,\,\,\,\,\,\,\,\,\,\,\,\,\,\, \textrm{For} \,\, G\mu = 10^{-18} \, .\nonumber
\end{eqnarray}
The ratio peaks at $G\mu\sim 10^{-18}$ when the crossover occurs, and drops off as $(G\mu)^{5/6}$ for smaller values.

In the case of Synchrotron radiation, we give an upper bound on the possible effect by using the upper bound on the primordial magnetic field obtained by the analysis of the CMB. The bound is \cite{CMBbound}
\begin{equation}
B(t_{rec}) \, < \, 10^{-9}  {\rm{Gauss}} \, ,
\end{equation} 
with appropriate redshifting to later times. Using this input, we find that the resulting upper bounds on the fraction $\mathcal{F}$ for these same values of $G\mu$ are
\begin{eqnarray}
\mathcal{F}_{Sync} \, &\sim& \, 10^{-23} \,\,\,\,\,\,\,\,\,\,\,\,\,\,\,\,\, \textrm{For} \,\, G\mu = 10^{-7}\\
\mathcal{F}_{Sync} \, &\sim& \, 10^{-6} \,\,\,\,\,\,\,\,\,\,\,\,\,\,\,\,\,\,\, \textrm{For} \,\, G\mu = 10^{-18} \, . \nonumber 
\end{eqnarray}
Once again, the maximal energy density in $21$-cm radiation peaks at the crossover value of $G\mu$ when evaluated at the time $t_{rec}$. 

The total energy density will be a superposition of the photons produced by Bremsstrahlung and by Synchrotron emission. Though Synchrotron seems to win by about $3-4$ orders of magnitude, we urge the reader to be cautious since we have used an upper bound on the primordial magnetic field and, in addition, a more detailed Synchrotron analysis (taking into account the beaming angle of the cusp emission, for example), may change the Synchrotron prediction considerably. 

In the Rayleigh-Jeans limit, we  find that the maximum induced temperature coming from a background of synchrotron photons sourced by cusp annihilations is
\begin{align}
T_{sync}^{strings} \, \sim \, 5 \cdot 10^{-3} \,\, \textrm{K} \, .
\end{align}

We can compute the differential brightness temperature at reionization from the expression
\begin{align}
\delta T_b = 27 x_{HI} \left(1-\frac{T_r}{T_s}\right) \sqrt{\frac{1+z}{10}\frac{0.15}{\Omega_mh^2}} \left(\frac{\Omega_b h^2}{0.023}\right) \,\, \textrm{mK}
\end{align}
Where $x_{HI}$ is the fraction of neutral hydrogen, and $T_s$ is the spin temperature of the hydrogen gas. We also define
\begin{align}
T_r = T_{CMB} + T^{strings}_{sync} + T^{strings}_{Brem}
\end{align}
As our new background temperature. We parametrize our enhancement over the standard CMB signal as
\begin{align}
\frac{\delta T_b}{\delta T_b^{CMB}} = 1+ \frac{T^{strings}_{sync} + T^{strings}_{Brem}}{T_{CMB} - T_s}
\end{align}
Where $\delta T_b^{CMB}$ is the differential brightness temperature in the case where the whole background is sourced by the CMB only. At reionization the spin and background temperatures can be calculated using the 21cmFAST code \cite{ARCADE2Gil} with $T_{CMB} \sim \mathcal{O}(10)$ K and $T_s \sim \mathcal{O}(1)$ K. This implies our maximum departure from standard CMB predictions is
\begin{align}
\frac{\delta T_b}{\delta T_b^{CMB}} \sim 1+ \mathcal{O}(10^{-4})
\end{align}
Thus, a precision of one part in $10^4$ is required by future experiments to constrain even the strongest of our signals. Currently, the only global measurement of this signal is from the EDGES collaboration, who are reporting a signal twice as strong as expected from CMB considerations alone ($\delta T_b^{EDGES}/\delta T_b^{CMB} \sim 2$). Our mechanism cannot explain this EDGES signal, though it could still have a small effect on the depth of the absorption trough that could be constrained if precision measurements of this global $21$-cm signal are achieved.

\section{Conclusions and Discussion}

In this work, we have considered the effects of cusp annihilations of cosmic strings on the radio  frequency radiation background present at reionization. We have utilized the fact that QCD jets emanating from cusp annihilations will produce a cascade of neutral and charged pions, which will quickly decay into electrons. These electrons interact with the dilute gas of hydrogen that exists between recombination and reionization, inducing a soft photon spectrum via Bremsstrahlung and Synchrotron radiation.

We have computed number density and energy density distributions for both of these processes. These computations are valid for any $E_{\gamma} < m_{\pi}$ and any time between recombination and reionization. For our purposes, we computed the additional contribution to the background radiation temperature in $21$-cm at reionization, in the hopes of using the upper bound on the global $21 {\rm{cm}}$ absorption signal at times of reionization to constrain the cosmic string tension. We found that the induced effect of cosmic strings increases as the string tension decreases, hence opening the possibility that a lower bound on the string tension could be establshed. However, since for low string tensions cusp annihilation becomes the dominant string decay mechanism and changes the string loop distribution, we found that the soft photon spectrum from strings peaks at a valuie of $G\mu \sim 10^{-18}$, and that even for that value of $G\mu$, the soft photons from cusp annihilation do not have a sufficient impact on the radiation temperature to yield an effect on the global $21 {\rm{cm}}$ signal which exceeds the upper bound set by the EDGES experiment (recall that we are using the value reported in \cite{EDGES} as an upper bound on the possible effect).

Concerning the Bremsstrahlung effect, we have not taken into account electrons produced from cosmic string cusp annihilations before the time of recombination. Taking these into account could potentially increase the cosmic string signal.

Concerning the effect due to Synchrotron radiation, we should note that the computation of the emission spectrum uses a simplifying assumption that the cusp emission is isotropic, which is not the case. A more detailed analysis taking the beaming effect into account might change the amplitude of $d\rho/dE$.

As well, backreaction on the cusp may also reduce any observational signal. Gravitational backreaction is capable of reducing the cusp region, yielding a decrease in the emitted energy off of a cusp \cite{OlumBackReaction}.

As a conclusion, we find that cosmic string cusp annihilation provides too weak of a soft photon flux to lead to constraints on the parameter space of the theory. We have, however, neglected soft photons produced directly from the jets. At the current time we do not know how to predict the magnitude of this effect.

\section*{Acknowledgements}

The research at McGill is supported in part by funds from NSERC and from the Canada Research Chair program. BC is supported in part by an MSI fellowship. We are grateful to Oscar Hernandez for stimulating discussions.


\begin{thebibliography}{99}

\bibitem{CSrevs}
A. Vilenkin and E.P.S. Shellard, \textit{Cosmic Strings and other
Topological Defects} (Cambridge Univ. Press, Cambridge, 1994);\\
M.~B.~Hindmarsh and T.~W.~B.~Kibble, 
``Cosmic strings'', 
Rept.~Prog.~Phys.\ {\bf 58}, 477 (1995) 
[arXiv:hep-ph/9411342];\\
R.~H.~Brandenberger,
  ``Topological defects and structure formation'',
  Int.\ J.\ Mod.\ Phys.\ A {\bf 9}, 2117 (1994)
  [arXiv:astro-ph/9310041].
  
\bibitem{RHBrev}
R.~H.~Brandenberger,
  ``Probing Particle Physics from Top Down with Cosmic Strings'',
  Universe {\bf 1}, no. 4, 6 (2013)
  [arXiv:1401.4619 [astro-ph.CO]].
    
\bibitem{stringsLSS}
A.~Vilenkin,
  ``Cosmological Density Fluctuations Produced by Vacuum Strings'',
  Phys.\ Rev.\ Lett.\  {\bf 46}, 1169 (1981)
  Erratum: [Phys.\ Rev.\ Lett.\  {\bf 46}, 1496 (1981)].
  doi:10.1103/PhysRevLett.46.1169, 10.1103/PhysRevLett.46.1496; \\
N.~Turok and R.~H.~Brandenberger,
  ``Cosmic Strings And The Formation Of Galaxies And Clusters Of Galaxies'',
  Phys.\ Rev.\ D {\bf 33}, 2175 (1986);\\
H. Sato, ``Galaxy Formation by Cosmic Strings'',
  Prog. Theor. Phys.\  {\bf 75}, 1342 (1986);\\
A. Stebbins, ``Cosmic Strings and Cold Matter'',
  Ap. J. (Lett.) {\bf 303}, L21 (1986).

\bibitem{noacoustic}
J.~Magueijo, A.~Albrecht, D.~Coulson and P.~Ferreira,
  ``Doppler peaks from active perturbations'',
  Phys.\ Rev.\ Lett.\  {\bf 76}, 2617 (1996)
  [arXiv:astro-ph/9511042];\\
U.~L.~Pen, U.~Seljak and N.~Turok,
  ``Power spectra in global defect theories of cosmic structure formation'',
  Phys.\ Rev.\ Lett.\  {\bf 79}, 1611 (1997)
  [arXiv:astro-ph/9704165];\\
L.~Perivolaropoulos,
  ``Spectral Analysis Of Microwave Background Perturbations Induced By Cosmic
  Strings'',
  Astrophys.\ J.\  {\bf 451}, 429 (1995)
  [arXiv:astro-ph/9402024].
 
\bibitem{CMBconstraints}
T.~Charnock, A.~Avgoustidis, E.~J.~Copeland and A.~Moss,
  ``CMB Constraints on Cosmic Strings and Superstrings'',
  arXiv:1603.01275 [astro-ph.CO];\\
C.~Dvorkin, M.~Wyman and W.~Hu,
  ``Cosmic String constraints from WMAP and the South Pole Telescope'',
  Phys.\ Rev.\ D {\bf 84}, 123519 (2011)
  [arXiv:1109.4947 [astro-ph.CO]];\\
   P.~A.~R.~Ade {\it et al.}  [Planck Collaboration],
  ``Planck 2013 results. XXV. Searches for cosmic strings and other topological defects'',
  Astron.\ Astrophys.\  {\bf 571}, A25 (2014)
  [arXiv:1303.5085 [astro-ph.CO]].
 
\bibitem{KS}
  N.~Kaiser and A.~Stebbins,
  ``Microwave Anisotropy Due To Cosmic Strings'',
  Nature {\bf 310}, 391 (1984);\\
   R.~Moessner, L.~Perivolaropoulos and R.~H.~Brandenberger,
  ``A Cosmic string specific signature on the cosmic microwave background,''
  Astrophys.\ J.\  {\bf 425}, 365 (1994)
  [astro-ph/9310001].
  
 \bibitem{Danos}
 R.~J.~Danos and R.~H.~Brandenberger,
  ``Canny Algorithm, Cosmic Strings and the Cosmic Microwave Background'',
  Int.\ J.\ Mod.\ Phys.\ D {\bf 19}, 183 (2010)
  [arXiv:0811.2004 [astro-ph]];\\
S.~Amsel, J.~Berger and R.~H.~Brandenberger,
  ``Detecting Cosmic Strings in the CMB with the Canny Algorithm'',
  JCAP {\bf 0804}, 015 (2008)
  [arXiv:0709.0982 [astro-ph]];\\
A.~Stewart and R.~Brandenberger,
  ``Edge Detection, Cosmic Strings and the South Pole Telescope'',
 JCAP {\bf 0902}, 009 (2009);
  [arXiv:0809.865 [astro-ph]];\\
L.~Hergt, A.~Amara, R.~Brandenberger, T.~Kacprzak and A.~Refregier,
  ``Searching for Cosmic Strings in CMB Anisotropy Maps using Wavelets and Curvelets,''
  arXiv:1608.00004 [astro-ph.CO];\\
  J.~D.~McEwen, S.~M.~Feeney, H.~V.~Peiris, Y.~Wiaux, C.~Ringeval and F.~R.~Bouchet,
  ``Wavelet-Bayesian inference of cosmic strings embedded in the cosmic microwave background,''
  arXiv:1611.10347 [astro-ph.IM];\\
 R.~Ciuca and O.~F.~Hernandez,
  ``A Bayesian Framework for Cosmic String Searches in CMB Maps,''
  JCAP {\bf 1708}, no. 08, 028 (2017)
  doi:10.1088/1475-7516/2017/08/028
  [arXiv:1706.04131 [astro-ph.CO]];\\
 R.~Ciuca, O.~F.~Hernandez and M.~Wolman,
  ``A Convolutional Neural Network For Cosmic String Detection in CMB Temperature Maps,''
  arXiv:1708.08878 [astro-ph.CO].

\bibitem{wakes}
J.~Silk and A.~Vilenkin,
  ``Cosmic Strings And Galaxy Formation'',
  Phys.\ Rev.\ Lett.\  {\bf 53}, 1700 (1984);\\
 M.~J.~Rees, 
``Baryon concentrations in string wakes at $z\gtrsim 200$: implications for galaxy formation and large-scale structure'',
 Mon.\ Not.\ Roy.\ Astron.\ Soc.\  {\bf 222}, 27 (1986);\\
T.~Vachaspati,
  ``Cosmic Strings and the Large-Scale Structure of the Universe'',
  Phys.\ Rev.\ Lett.\  {\bf 57}, 1655 (1986);\\
A.~Stebbins, S.~Veeraraghavan, R.~H.~Brandenberger, J.~Silk and N.~Turok,
  ``Cosmic String Wakes'',
  Astrophys.\ J.\  {\bf 322}, 1 (1987);\\
  D.~Cunha, J.~Harnois-Deraps, R.~Brandenberger, A.~Amara and A.~Refregier,
  ``Dark Matter Distribution Induced by a Cosmic String Wake in the Nonlinear Regime,''
  arXiv:1804.00083 [astro-ph.CO];\\
 S.~Laliberte, R.~Brandenberger and D.~C.~N.~da Cunha,
  ``Cosmic String Wake Detection using 3D Ridgelet Transformations,''
  arXiv:1807.09820 [astro-ph.CO].
     
\bibitem{GC}
 L.~Lin, S.~Yamanouchi and R.~Brandenberger,
  ``Effects of Cosmic String Velocities and the Origin of Globular Clusters,''
  JCAP {\bf 1512}, no. 12, 004 (2015)
  doi:10.1088/1475-7516/2015/12/004
  [arXiv:1508.02784 [astro-ph.CO]];\\
  A.~Barton, R.~H.~Brandenberger and L.~Lin,
  ``Cosmic Strings and the Origin of Globular Clusters,''
  JCAP {\bf 1506}, no. 06, 022 (2015)
  doi:10.1088/1475-7516/2015/06/022
  [arXiv:1502.07301 [astro-ph.CO]].
  
\bibitem{bryceFRB}
R.~Brandenberger, B.~Cyr and A.~V.~Iyer,
  ``Fast Radio Bursts from the Decay of Cosmic String Cusps,''
  arXiv:1707.02397 [astro-ph.CO].
  
\bibitem{superconductingFRB}
T.~Vachaspati,
  ``Cosmic Sparks from Superconducting Strings,''
  Phys.\ Rev.\ Lett.\  {\bf 101}, 141301 (2008)
  doi:10.1103/PhysRevLett.101.141301
  [arXiv:0802.0711 [astro-ph]];\\
  L.~V.~Zadorozhna and B.~I.~Hnatyk,
  ``Electromagnetic emission bursts from the near-cusp regions of superconducting cosmic strings,''
  Ukr.\ J.\ Phys.\  {\bf 54}, 1149 (2009);\\
Y.~F.~Cai, E.~Sabancilar and T.~Vachaspati,
  ``Radio bursts from superconducting strings,''
  Phys.\ Rev.\ D {\bf 85}, 023530 (2012)
  doi:10.1103/PhysRevD.85.023530
  [arXiv:1110.1631 [astro-ph.CO]];\\
 Y.~F.~Cai, E.~Sabancilar, D.~A.~Steer and T.~Vachaspati,
  ``Radio Broadcasts from Superconducting Strings,''
  Phys.\ Rev.\ D {\bf 86}, 043521 (2012)
  doi:10.1103/PhysRevD.86.043521
  [arXiv:1205.3170 [astro-ph.CO]];\\
 J.~Ye, K.~Wang and Y.~F.~Cai,
  ``Superconducting cosmic strings as sources of cosmological fast radio bursts,''
  Eur.\ Phys.\ J.\ C {\bf 77}, no. 11, 720 (2017)
  doi:10.1140/epjc/s10052-017-5319-2
  [arXiv:1705.10956 [astro-ph.HE]];\\
 Y.~W.~Yu, K.~S.~Cheng, G.~Shiu and H.~Tye,
  ``Implications of fast radio bursts for superconducting cosmic strings,''
  JCAP {\bf 1411}, no. 11, 040 (2014)
  doi:10.1088/1475-7516/2014/11/040
  [arXiv:1409.5516 [astro-ph.HE]];\\
 L.~V.~Zadorozhna,
  ``Fast radio bursts as electromagnetic radiation fro cusps on 
  superconducting cosmic strings'',
  Adv. in Astronomy and Space Physics {\bf 5}, 43 (2015).
      
\bibitem{SMBH}
S.~F.~Bramberger, R.~H.~Brandenberger, P.~Jreidini and J.~Quintin,
  ``Cosmic String Loops as the Seeds of Super-Massive Black Holes,''
  JCAP {\bf 1506}, no. 06, 007 (2015)
  doi:10.1088/1475-7516/2015/06/007
  [arXiv:1503.02317 [astro-ph.CO]].
  
\bibitem{CSsimuls}
A.~Albrecht and N.~Turok,
  ``Evolution Of Cosmic Strings'',
  Phys.\ Rev.\ Lett.\  {\bf 54}, 1868 (1985);\\
D.~P.~Bennett and F.~R.~Bouchet,
  ``Evidence For A Scaling Solution In Cosmic String Evolution'',
  Phys.\ Rev.\ Lett.\  {\bf 60}, 257 (1988);\\
B.~Allen and E.~P.~S.~Shellard,
  ``Cosmic String Evolution: A Numerical Simulation'',
  Phys.\ Rev.\ Lett.\  {\bf 64}, 119 (1990);\\
C.~Ringeval, M.~Sakellariadou and F.~Bouchet,
  ``Cosmological evolution of cosmic string loops'',
  JCAP {\bf 0702}, 023 (2007)
  [arXiv:astro-ph/0511646];\\
V.~Vanchurin, K.~D.~Olum and A.~Vilenkin,
  ``Scaling of cosmic string loops'',
  Phys.\ Rev.\  D {\bf 74}, 063527 (2006)
  [arXiv:gr-qc/0511159];\\
L.~Lorenz, C.~Ringeval and M.~Sakellariadou,
  ``Cosmic string loop distribution on all length scales and at any redshift'',
  JCAP {\bf 1010}, 003 (2010)
  [arXiv:1006.0931 [astro-ph.CO]];\\
J.~J.~Blanco-Pillado, K.~D.~Olum and B.~Shlaer,
  ``Large parallel cosmic string simulations: New results on loop production'',
  Phys.\ Rev.\ D {\bf 83}, 083514 (2011)
  [arXiv:1101.5173 [astro-ph.CO]];\\
J.~J.~Blanco-Pillado, K.~D.~Olum and B.~Shlaer,
  ``The number of cosmic string loops'',
  Phys.\ Rev.\ D {\bf 89}, no. 2, 023512 (2014)
  [arXiv:1309.6637 [astro-ph.CO]].

\bibitem{PTAconstraints}
Z.~Arzoumanian {\it et al.} [NANOGrav Collaboration],
  ``The NANOGrav Nine-year Data Set: Limits on the Isotropic Stochastic Gravitational Wave Background,''
  Astrophys.\ J.\  {\bf 821}, no. 1, 13 (2016)
  doi:10.3847/0004-637X/821/1/13
  [arXiv:1508.03024 [astro-ph.GA]].

\bibitem{ARCADE2}
D.J. ~Fixsen {\it et al.},
  ``ARCADE 2 Measurement of the Absolute Sky Brightness at 3-90 GHz''
  Astrophys.\ J\. {\bf 734}, no. 1, (2011)
  doi:10.1088/0004-637X/734/1/5

\bibitem{ARCADE2Gil}
C.~Feng and G.~Holder,
  ``Enhanced global signal of neutral hydrogen due to excess radiation at cosmic dawn,''
  Astrophys.\ J.\  {\bf 858}, no. 2, L17 (2018)
  doi:10.3847/2041-8213/aac0fe
  [arXiv:1802.07432 [astro-ph.CO]].
  
\bibitem{Furl}
S.~Furlanetto, S.~P.~Oh and F.~Briggs,
  ``Cosmology at Low Frequencies: The 21 cm Transition and the High-Redshift Universe,''
  Phys.\ Rept.\  {\bf 433}, 181 (2006)
  doi:10.1016/j.physrep.2006.08.002
  [astro-ph/0608032].
  
\bibitem{EDGES}
J. D. Bowman, A. E. E. Rogers, R. A. Monsalve, T. J. Mozdzen and N. Mahesh, 
Nature 555, no. 7694, 67 (2018);\\
R.~A.~Monsalve, B.~Greig, J.~D.~Bowman, A.~Mesinger, A.~E.~E.~Rogers, T.~J.~Mozdzen, N.~S.~Kern and N.~Mahesh,
  ``Results from EDGES High-Band: II. Constraints on Parameters of Early Galaxies,''
  Astrophys.\ J.\  {\bf 863}, no. 1, 11 (2018)
  doi:10.3847/1538-4357/aace54
  [arXiv:1806.07774 [astro-ph.CO]].

\bibitem{EDGEScritic}
R.~Hills, G.~Kulkarni, P.~D.~Meerburg and E.~Puchwein,
  ``Concerns about Modelling of Foregrounds and the 21-cm Signal in EDGES data,''
  arXiv:1805.01421 [astro-ph.CO].
  
\bibitem{Oscar}
O.~F.~Hernandez,
  ``Wouthuysen-Field absorption trough in cosmic string wakes,''
  Phys.\ Rev.\ D {\bf 90}, no. 12, 123504 (2014)
  doi:10.1103/PhysRevD.90.123504
  [arXiv:1403.7522 [astro-ph.CO]].

\bibitem{milliDark}
A.~Fialkov, R.~Barkana and A.~Cohen,
  ``Constraining Baryon--Dark Matter Scattering with the Cosmic Dawn 21-cm Signal,''
  Phys.\ Rev.\ Lett.\  {\bf 121}, 011101 (2018)
  doi:10.1103/PhysRevLett.121.011101
  [arXiv:1802.10577 [astro-ph.CO]];\\
R.~Barkana, N.~J.~Outmezguine, D.~Redigolo and T.~Volansky,
  ``Signs of Dark Matter at 21-cm?,''
  arXiv:1803.03091 [hep-ph];\\
S.~Fraser {\it et al.},
  ``The EDGES 21 cm Anomaly and Properties of Dark Matter,''
  arXiv:1803.03245 [hep-ph].
  
\bibitem{Kibble}
T.~W.~B.~Kibble,
  ``Phase Transitions In The Early Universe'',
  Acta Phys.\ Polon.\  B {\bf 13}, 723 (1982);\\
  T.~W.~B.~Kibble,
  ``Some Implications Of A Cosmological Phase Transition'',
  Phys.\ Rept.\  {\bf 67}, 183 (1980).
  
\bibitem{scalingSolution}
E.~J.~Copeland, T.~W.~B.~Kibble and D.~Austin,
  ``Scaling solutions in cosmic string networks,''
  Phys.\ Rev.\ D {\bf 45}, 1000 (1992).
  doi:10.1103/PhysRevD.45.R1000;\\
L.~Perivolaropoulos,
  ``COBE versus cosmic strings: An Analytical model,''
  Phys.\ Lett.\ B {\bf 298}, 305 (1993)
  doi:10.1016/0370-2693(93)91825-8
  [hep-ph/9208247];\\
  D.~Austin, E.~J.~Copeland and T.~W.~B.~Kibble,
  ``Evolution of cosmic string configurations,''
  Phys.\ Rev.\ D {\bf 48}, 5594 (1993)
  doi:10.1103/PhysRevD.48.5594
  [hep-ph/9307325].
  
\bibitem{Hindmarsh}
M.~Hindmarsh, J.~Lizarraga, J.~Urrestilla, D.~Daverio and M.~Kunz,
  ``Scaling from gauge and scalar radiation in Abelian Higgs string networks,''
  arXiv:1703.06696 [astro-ph.CO].
  
\bibitem{KibbleTurok}
T.~W.~B.~Kibble and N.~Turok,
  ``Selfintersection of Cosmic Strings,''
  Phys.\ Lett.\  {\bf 116B}, 141 (1982).
  doi:10.1016/0370-2693(82)90993-5
  
\bibitem{cuspPartProduction}
 R.~H.~Brandenberger,
  ``On the Decay of Cosmic String Loops,''
  Nucl.\ Phys.\ B {\bf 293}, 812 (1987).
  doi:10.1016/0550-3213(87)90092-7

\bibitem{robertGamma}
J.~H.~MacGibbon and R.~H.~Brandenberger,
  ``Gamma-ray signatures from ordinary cosmic strings,''
  Phys.\ Rev.\ D {\bf 47}, 2283 (1993)
  doi:10.1103/PhysRevD.47.2283
  [astro-ph/9206003].

\bibitem{robertNeutrino}
J.~H.~MacGibbon and R.~H.~Brandenberger,
  ``High-energy neutrino flux from ordinary cosmic strings,''
  Nucl.\ Phys.\ B {\bf 331}, 153 (1990);\\
  doi:10.1016/0550-3213(90)90020-E
U.~F.~Wichoski, J.~H.~MacGibbon and R.~H.~Brandenberger,
  ``High-energy neutrinos, photons and cosmic ray fluxes from VHS cosmic strings,''
  Phys.\ Rev.\ D {\bf 65}, 063005 (2002)
  doi:10.1103/PhysRevD.65.063005
  [hep-ph/9805419].
  
\bibitem{olumCusp}
J.~J.~Blanco-Pillado and K.~D.~Olum,
  ``The Form of cosmic string cusps,''
  Phys.\ Rev.\ D {\bf 59}, 063508 (1999)
  doi:10.1103/PhysRevD.59.063508
  [gr-qc/9810005].

\bibitem{widthRef}
H.~B.~Nielsen and P.~Olesen,
  ``Vortex Line Models for Dual Strings,''
  Nucl.\ Phys.\ B {\bf 61}, 45 (1973).
  doi:10.1016/0550-3213(73)90350-7
  
\bibitem{blumenthal}
G.~R.~Blumenthal and R.~J.~Gould,
  ``Bremsstrahlung, synchrotron radiation, and compton scattering of high-energy electrons traversing dilute gases,''
  Rev.\ Mod.\ Phys.\  {\bf 42}, 237 (1970).
  doi:10.1103/RevModPhys.42.237

\bibitem{robertSpectral}
J.~H.~MacGibbon and R.~H.~Brandenberger,
  ``High-energy neutrino flux from ordinary cosmic strings,''
  Nucl.\ Phys.\ B {\bf 331}, 153 (1990).
  doi:10.1016/0550-3213(90)90020-E

 \bibitem{CSgrav}
   T.~Vachaspati and A.~Vilenkin,
  ``Gravitational Radiation from Cosmic Strings,''
  Phys.\ Rev.\ D {\bf 31}, 3052 (1985).
  
\bibitem{Holder}
  M.~Kaplinghat, M.~Chu, Z.~Haiman, G.~Holder, L.~Knox and C.~Skordis,
  ``Probing the reionization history of the universe using the cosmic microwave background polarization,''
  Astrophys.\ J.\  {\bf 583}, 24 (2003)
  doi:10.1086/344927
  [astro-ph/0207591].
  
\bibitem{CMBbound}
  A. ~Zucca, Y. ~Li, L. ~Pogosian,
  ``Constraints on primordial magnetic fields from Planck data combined with the South Pole Telescope CMB $B$-mode polarization measurements,''
  Phys.\ Rev.\ D {\bf 95}, 063506 (2017)
  doi:10.1103/PhysRevD.95.063506
  [astro-ph/1611.00757].
  

\bibitem{OlumBackReaction}
  J.J. ~Blanco-Pillado, K.D. ~Olum, J.M. ~Wachter,
  ``Gravitational back-reaction near cosmic string kinks and cusps,''
  [gr-qc/1808.08254]
  
\end{thebibliography}
\end{document}